\documentclass[runningheads]{llncs}
\usepackage{graphicx}
\usepackage{booktabs}
\usepackage{multirow}
\usepackage{amssymb}
\usepackage{adjustbox}
\usepackage{amsmath}
\usepackage{amssymb}
\usepackage{bm}
\usepackage{cite}
\usepackage{subcaption}
\usepackage[misc]{ifsym} 
\usepackage{hyperref}
\hypersetup{
    colorlinks=true,
    linkcolor=blue,
    filecolor=magenta,      
    urlcolor=cyan,
    pdftitle={Overleaf Example},
    pdfpagemode=FullScreen,
    }

\begin{document}
\title{CAS-Net: Conditional Atlas Generation and Brain Segmentation for Fetal MRI}

\titlerunning{CAS-Net: Conditional Atlas Generation and Brain Segmentation}
%
\author{Liu Li\inst{1}$^{(\textrm{\Letter})}$ \and
Qiang Ma\inst{1} \and
Matthew Sinclair\inst{1} \and
Antonios Makropoulos\inst{1} \and
Joseph Hajnal\inst{2} \and \\
A. David Edwards\inst{2} \and
Bernhard Kainz \inst{1,2,3} \and
Daniel Rueckert \inst{1,4} \and
Amir Alansary \inst{1}
}
%
\authorrunning{Liu et al.}
%
\institute{BioMedIA Group, Department of Computing, Imperial College London, UK  \email{liu.li20@imperial.ac.uk} \and
King’s College London, UK, 
$^3$ FAU Erlangen-N\"urnberg, Germany,  \\ $^4$ Technical University of Munich, Germany}

%
\maketitle              


\begin{abstract}
Fetal Magnetic Resonance Imaging (MRI) is used in prenatal diagnosis and to assess early brain development.
Accurate segmentation of the different brain tissues is a vital step in several brain analysis tasks, such as cortical surface reconstruction and tissue thickness measurements.
Fetal MRI scans, however, are prone to motion artifacts that can affect the correctness of both manual and automatic segmentation techniques. 
In this paper, we propose a novel network structure that can simultaneously generate conditional atlases and predict brain tissue segmentation, called CAS-Net. 
The conditional atlases provide anatomical priors that can constrain the segmentation connectivity, despite the heterogeneity of intensity values caused by motion or partial volume effects.
The proposed method is trained and evaluated on 253 subjects from the developing Human Connectome Project (dHCP).
The results demonstrate that the proposed method can generate conditional age-specific atlas with sharp boundary and shape variance. It also segments  multi-category brain tissues for fetal MRI with a high overall Dice similarity coefficient (DSC) of $85.2\%$ for the selected 9 tissue labels. 
\end{abstract}


\section{Introduction}

The perinatal period is an important time for the study of human brain development. 
Cellular connections start to form across the brain, and the cerebral cortex becomes more complex. 
These brain structure developments are closely related to the formation of human cognitive functions \cite{casey2000structural}. 
Magnetic Resonance Imaging (MRI) plays an essential role in fetal diagnosis and for studying neurodevelopment, as it can capture different brain tissues in detail compared to conventional fetal Ultrasound~\cite{ertl2002fetal}.

Brain tissue segmentation is important for the quantitative evaluation of the cortical development, and is a vital step for standard surface reconstruction pipelines \cite{makropoulos2018developing}.
However, during the scanning process, the fetus is moving and not sedated and the mother breathes normally, which can produce motion artifacts. 
Recent works for super-resolution and motion correction in fetal MRI \cite{alansary2017pvr,ebner2020automated,kuklisova2012reconstruction} can reconstruct the scanned image with less motion artifacts between the slices; nevertheless, in-plane motion can remain.

Other limitations are partial volume effects and lower signal-to-noise-ratio caused by the small size of the brain.
Furthermore, the structure of the fetal brain has a large shape variance because of the rapid development during the perinatal period. 
This can hinder the learning of automatic segmentation models, especially using a limited number of training subjects from a wide age range.

In order to address problems caused by poor image quality and inaccurate training labels, we propose a novel architecture that learns to predict the segmentation maps and a conditional atlas simultaneously in an end-to-end pipeline. 
The atlas enables the model to learn anatomical priors without depending solely on the intensity values of the input image. This can improve the segmentation performance especially if there is no gold standard label for training due to the poor image quality.


\subsection{Related work}
Traditionally, the two tasks of atlas generation and tissue segmentation are learned separately. 
For tissue segmentation, atlas-based segmentation results are obtained by warping the atlas label maps to the target image with a restrained deformation field \cite{cabezas2011review, makropoulos2014automatic}. 
Generating an atlas with higher similarity to the target image will ease the calculation for the deformation field thus improving the segmentation performance. For atlas generation, the atlas is usually generated by registering and aggregating all the images to the same atlas coordinate space. 
The performance of these methods largely relies on the initial similarity of the atlas and the testing samples \cite{makropoulos2014automatic}. 
Since atlas-based approaches require expensive registration optimization, the processing time may take hours to days for a single subject. 

With the evolution of deep learning methods, UNet-based~\cite{ronneberger2015u} architectures have defined state-of-the-art performance for segmentation \cite{isensee2018nnu,zhou2018unet++,alom2018recurrent}, including fetal brain scans \cite{khalili2019automatic,dou2020deep,payette2020automatic}. 
Such methods can be sensitive to the consistency of the intensity distribution of the input images and the accuracy of the segmentation label maps used for training.
Fetit et al. \cite{fetit2020deep} proposed a segmentation pipeline for neonatal CGM that utilized the silver labels from the DrawEM method \cite{makropoulos2014automatic}, and further used a small part of the manually labeled slices to refine the segmentation results. 

Dalca et al. \cite{dalca2018anatomical} trained an additional auto-encoding variational network to encode the anatomical prior in a decoder. However, the anatomical priors are implicitly encoded in these methods, which can be difficult to interpret manually.
Recently, Sinclair et al. \cite{sinclair2020atlas} proposed a network that jointly learns segmentation and registration, and atlas construction for cardiac and brain scans.
It explicitly learns the shape prior based on a structure-guided image registration \cite{lee2019image}.

\noindent\textbf{Contributions:} 
Inspired by previous works \cite{dalca2018anatomical,fetit2020deep,sinclair2020atlas}, we propose a novel 
segmentation network that utilizes anatomical priors conditioned on an age-specific 4D atlas to address the challenges caused by noisy labels and bad image quality.
The main contributions of this work include:
\begin{itemize}
    \item A novel multi-task network for learning end-to-end tissue segmentation and conditional atlas generation simultaneously, called CAS-Net.
    \item An age-conditioned 4D atlas construction that is constrained by the smoothness and continuity of the diffeomorphic transformation field.
    \item A detailed evaluation of the quality of the spatio-temporal atlas and the performance of the segmentation, whereas the proposed CAS-Net outperforms other baseline methods with an overall $85.2\%$ dice score.
\end{itemize}





\section{Method}


\begin{figure}[htbp]
\centering
    \hfill
    \begin{subfigure}[b]{0.49\textwidth}
        \centering
        \includegraphics[width=1.0\linewidth]{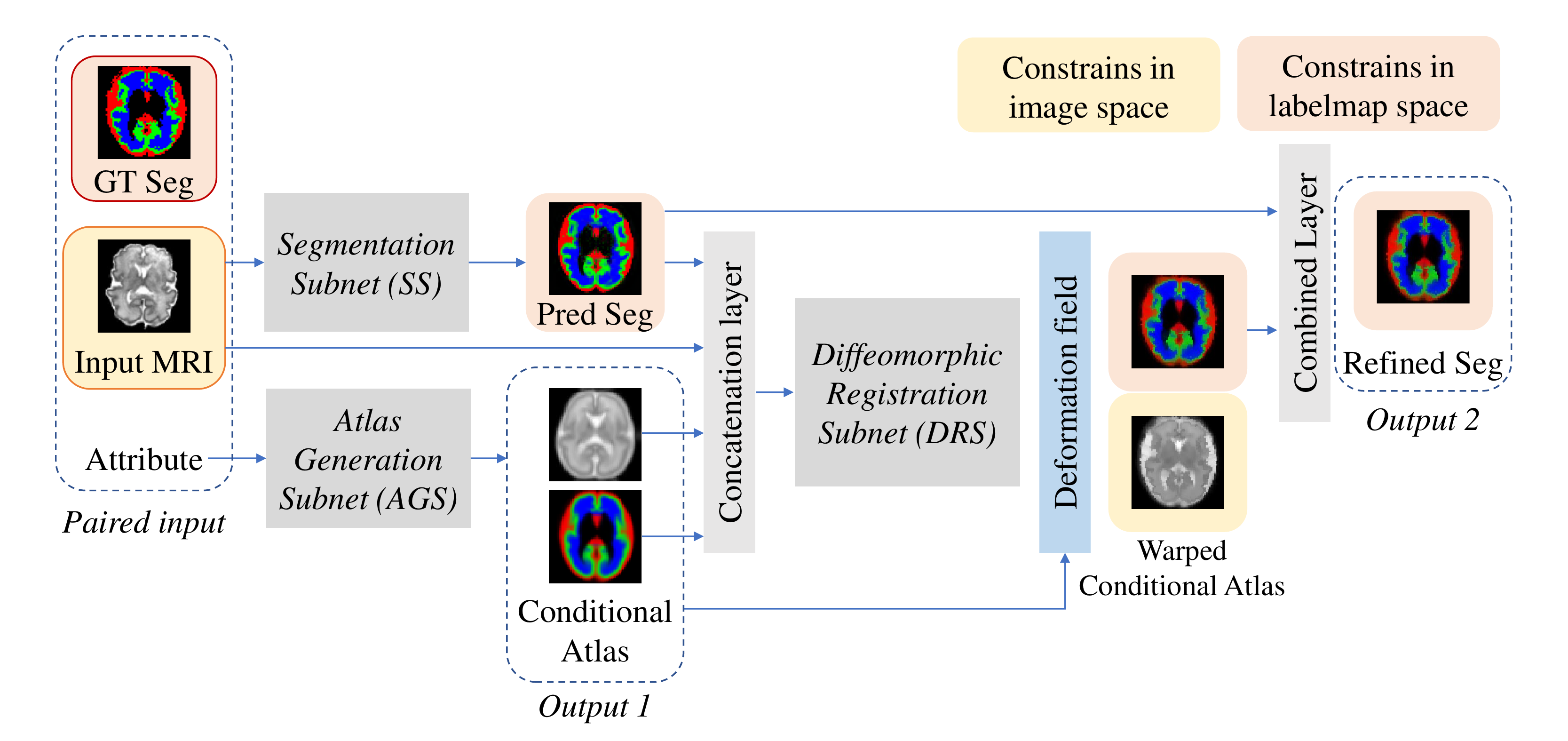}
        \caption{CAS-Net}
        \label{fig:overviewA}
    \end{subfigure}
    \hfill
    \begin{subfigure}[b]{0.49\textwidth}
        \centering
        \includegraphics[width=1.0\linewidth]{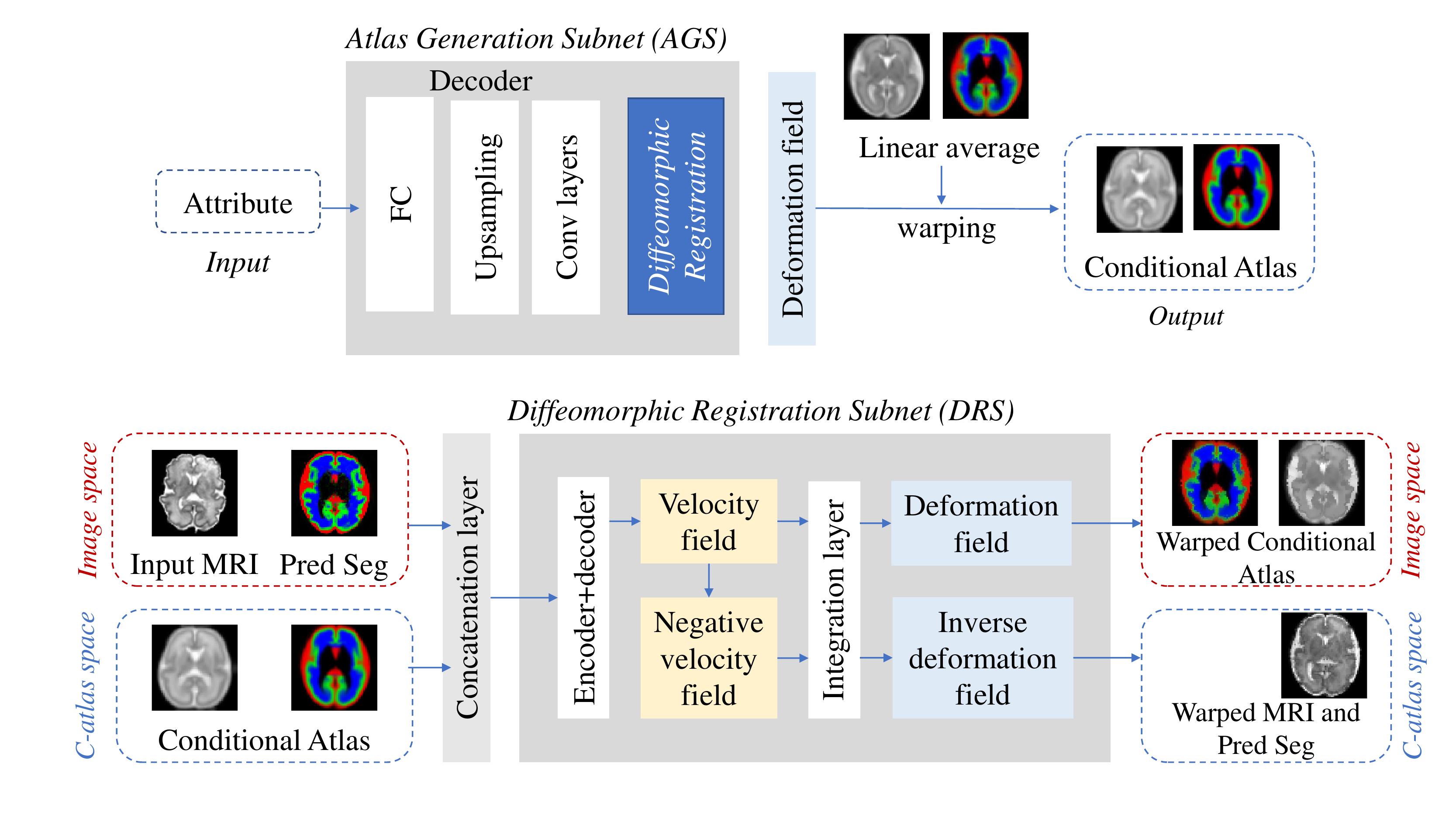}
        \caption{AGS (top) and DRS (bottom)}
        \label{fig:overviewB}
    \end{subfigure}
    \hfill
    \caption{\footnotesize The main architecture of the proposed CAS-Net for brain tissue segmentation and conditional atlas generation (a). The detailed architecture of the atlas generation (AGS) and diffeomorphic registration (DRS) subnets (b).}
\label{fig:overview}
\end{figure}

CAS-Net learns end-to-end the conditional atlas generation and tissue segmentation, whereas the age-specific atlas can be considered as an anatomical prior for the segmentation task. 
It consists of three main sub-networks: segmentation subnet (SS), atlas generation subnet (AGS), and diffeomorphic registration subnet (DRS), see Fig.~\ref{fig:overview}. 

The input of the CAS-Net, Fig. \ref{fig:overviewA}, is the paired MR image, segmentation labelmap and the conditional attribute (\emph{e.g.}, gestational age) from the training set $\mathcal{D}=\{\bm{I}_i, \bm{S}_i, \bm{a}_i\}$, where $i$ is the index of training samples. Note that the MRI and its labelmap are in the size of $l\times w \times h \times 1$ and $l\times w \times h \times c$, where $l, w$ and $h$ are the length, width and height of the 3D volume, and $c$ is the number of anatomical labels in the segmentation maps. 
In our setting, the conditional attribute is the gestational age (GA), which is used to generate an age-specific atlas for studying the development of brain structure. However, other attributes such as sex or pathologies camayn can also be used for conditional atlas generation.

The input of the AGS, Fig.~\ref{fig:overviewB}, are the GA attribute $\bm{a}_i$, the global averaging atlas image $\bm A_{\mathrm{g}}$ and its labelmap $\bm A^{\mathrm{s}}_{\mathrm{g}}$, and the output are the conditional atlas image $\bm A$ and its labelmap $\bm{A}^{\mathrm{s}}$. At the same time, the segmentation subnet takes the MR image $\bm{I}_i$ as an input, and predicts the multi-label segmentation map $\bm{\hat{S}}_{i}$. 

As such, both $\bm{I}_i$ and $\bm{\hat{S}}_{i}$ are in the image space, and conditional atlas space ($\bm A$ and $\bm{A}^{\mathrm{s}}$).
Then, the four images ($\bm{I}_i$, $\bm{\hat{S}}_{i}$, $\bm A$ and $\bm{A}^{\mathrm{s}}$) are concatenated along the channel axis as the input for the diffeomorphic registration subnet.
Based on the concatenated input, a deformation field $\bm{\Phi}_i$ is predicted by the DRS to warp the conditional atlas to the image space, and output the atlas-based segmentation result $\bm{\hat{S}}^{\mathrm {a}}_{i}$.
Finally, $\bm{\hat{S}}_{i}$ and $\bm{\hat{S}}^{\mathrm {a}}_{i}$ are merged by a convolutional layer, i.e., a 3D convolutional layer with the kernal size of $1\times 1 \times 1$, to output the final refined segmentation results $\bm{\hat{S}}^{\mathrm {r}}_{i}$.


\noindent\textbf{Diffeomorphic Registration Subnet (DRS):}
The purpose of the DRS is to learn a deformation field that registers the conditional atlas to the target input image, so that the atlas label maps can also be warped and propagated to segment the target image.
Inspired by \cite{dalca2019learning, sinclair2020atlas}, we model the registration step using a diffeomorphic deformation parameterized by a stationary velocity field, which ensures a one-to-one mapping while preserving the topology of the brain structure \cite{ashburner2007fast}. 
Given that the diffeomorphic deformation is always invertible, the deformation between the conditional atlas and input image is also bidirectional.

Given a deformation field $\bm{\Phi}$, the process of warping the moving atlas can be formulated as $\bm{A} \circ \bm{\Phi}$, where $\circ$ is the operation of using the values from $\bm{A}$ and voxel locations from $\bm{\Phi}$ to calculate the output warped image.
Note that the 3D moving atlas and the output warped image are both of size of $l\times w \times h \times c$, where $c=1$ for the atlas image and $c=k$ for atlas labelmap, and the learnt deformation field $\bm{\Phi}$ is in the size of $l\times w \times h \times 3$.

The procedure for computing a diffeomorphic deformation can be modeled by an ordinary differential equation (ODE) parameterized with a stationary velocity field $\bm{V}$ \cite{ashburner2007fast}:
\begin{equation}
    \frac{\mathrm{d}\bm{\Phi}}{\mathrm{d}t}=\bm{V}(\bm{\Phi^{(t)}}).
\end{equation}
As shown in Fig.~\ref{fig:overviewB} (bottom), $\bm{V}$ is simulated by an encoder-decoder structure with the concatenated input.
Given the velocity field, the final deformation field $\bm{\Phi}^{(t=1)}$ is obtained by integrating the stationary velocity field $\bm{V}$ over unit time:
\begin{equation}
    \bm{\Phi}^{(t=1)} = \bm{\Phi}^{(t=0)} + \int_{t=0}^{t=1}\bm{V(\bm{\Phi}^{(t)})}\mathrm{d}t.  
\end{equation}
Here, $\bm{\Phi}^{(t=0)}$ is the initial condition of identity deformation, i.e., $\bm{X} \circ \bm{\Phi}_{t=0}=\bm{X}$, where $\bm{X}$ is a map with the size of $l\times w \times h \times c$.

In practice, the computation of the integration is approximated by the Euler methods with small time steps $h$, and is interpolated by the scaling and squaring layer in CAS-Net, similar to \cite{dalca2018unsupervised}. Specifically, the integration is recursively calculated with small time steps $h$ as: 
\begin{equation}
\begin{aligned}
    \bm{\Phi}^{(t+h)} = \bm{\Phi}^{(t)} + \int_{t}^{t+h}\bm{V(\bm{\Phi}^{(t)})}\mathrm{d}t 
    \approx \bm{\Phi}^{(t)} + h\bm{V(\bm{\Phi}^{(t)})} 
    = (x + h\bm{V}) \circ \bm{\Phi}^{(t)},
\end{aligned}
\end{equation}
with $x = \bm{\Phi}^{(0)} $. When the integration steps in the scaling and squaring layer is the power of 2, given the predicted velocity field $\bm V$, the final deformation field $\bm{\Phi}^{(1)}$ can be calculated iteratively through Euler integration.

Since the diffeomorphic registration is invertible, the image can be warped to the atlas space using the inverse deformation field $\bm{\Phi}_i^{-1}$, which is calculated by integrating the negative velocity field $\bm{-V}_i$, as shown in Fig.~\ref{fig:overviewB} (bottom).


\noindent\textbf{Atlas Generation Subnet (AGS):} The main task of the AGS is to generate an age specific deformation field that can warp the global average atlas to be age-specific.
Different to \cite{evan2020learning}, which only constructs the conditional atlas image by adding a displacement field, AGS predicts the deformation field, atlas image and labelmap simultaneously. The age-specific atlas labelmap is then used by the DRS to predict the atlas-based segmentation result.

The structure of this subnet is shown in Fig.~\ref{fig:overviewB} (top). The input attribute is first decoded from the low-dimensional attribute space $\bm{\mathrm{a}}$ to a high-dimensional feature space $\bm{Q}_i$. Specifically, the decoder consists of a fully connected layer, an upsampling layer and several convolutional layers.
In order to deal with the problem of the imbalanced distribution of GA, we divide the training samples into 4 age groups, and encode this attribute in a one-hot label (instead of directly encoding the GA as a continuous scalar).
Based on the decoder feature map $\bm{Q}_i$, we also model the deformation with a diffeomorphism parameterized by a velocity field, where $\bm{Q}_i$ is treated as the velocity field. 
Then, the age-specific $\bm{Q}_i$ is integrated by the scaling and squaring layer $\bm{\Psi}_i$. Consequently, the conditional atlas image and labelmap are constructed as: 
$\bm A = \bm A_{\mathrm{g}} \circ \bm{\Psi}_i,$ and 
$\bm A^{\mathrm{s}} = \bm A^{\mathrm{s}}_{\mathrm{g}} \circ \bm{\Psi}_i$.
Here, $\bm A_{\mathrm{g}}$ and $\bm A^{\mathrm{s}}_{\mathrm{g}}$ are the global atlas image and labelmap, respectively, which are initialized by averaging all the input images $\bm{I}_i$ and their corresponding labelmaps $\bm{S}_{i}$ in the training set.

In addition, since the deformation between global atlas space to conditional atlas space ($\bm{\Phi}_i$ and $\bm{\Phi}_i^{-1}$) and the deformation between conditional atlas space and input MRI space ($\bm{\Psi}_i$ and $\bm{\Psi}_i^{-1}$) are diffeomorphic, following \cite{sinclair2020atlas}, the global atlas and its labelmap in our CAS-Net are also updated at the end of each epoch to improve the segmentation performance using: 
$\bm A_{\mathrm{g}}^{(j)} = \frac{1}{N}\sum^{N}_{i=1} \bm{I}_i \circ \bm{\Phi}_i^{-1(j)} \circ \bm{\Psi}_i^{-1(j)},$
and 
$\bm A_{\mathrm{g}}^{\mathrm{s}(j)} = \frac{1}{N}\sum^{N}_{i=1} \bm{S}_i \circ \bm{\Phi}_i^{-1(j)} \circ \bm{\Psi}_i^{-1(j)}$.
Here, $j$ is the index of epoch, and $N$ is the number of the training samples.


\noindent\textbf{Loss Function:} 
In order to achieve end-to-end training, the training process is optimized using four loss terms, namely, segmentation loss $\mathcal{L}_{\mathrm{S}}$, registration loss $\mathcal{L}_{\mathrm{R}}$, combination loss $\mathcal{L}_{\mathrm{C}}$ and the regularization term $\mathcal{L}_{\mathrm{Reg}}$. 

The segmentation loss is a standard $L^2$-norm between the predicted $\bm{\hat{S}}_{i}$ and groundtruth labels $\bm{S}_i$, and used to update the parameters of the segmentation subnet.

The parameters of the AGS and DRS are supervised by $L_{\mathrm{R}}$ in both image and labelmap space as follows:
\begin{equation}
    \begin{aligned}
    \mathcal{L}_{\mathrm{R}} =& \lambda_{\mathrm{l}}\| \bm A^{\mathrm{s}} \circ \bm{\Phi}_i - \bm{S}_i \|_{2} +
    \lambda_{\mathrm{i}}\| \bm A \circ \bm{\Phi}_i - \bm{I}_i\|_{2} \\
    =&\lambda_{\mathrm{l}}\| \bm A^{\mathrm{s}}_{\mathrm{g}} \circ \bm{\Psi}_i \circ \bm{\Phi}_i - \bm{S}_i \|_{2} +
    \lambda_{\mathrm{i}}\| \bm A_{\mathrm{g}} \circ \bm{\Psi}_i \circ \bm{\Phi}_i - \bm{I}_i\|_{2}.
    \end{aligned}
\end{equation}
Here, $\lambda_{\mathrm{i}}$ and $\lambda_{\mathrm{l}}$ are the weights for image and labelmap space loss. The first term is beneficial for the generated atlas quality, while the second term contributes to the accuracy of the segmentation.
At the beginning of the training, $\lambda_{\mathrm{i}}$ has higher values in order to learn an accurate conditional atlas, and later $\lambda_{\mathrm{l}}$ is increased for a better segmentation performance. 

The combination loss ($\mathcal{L}_{\mathrm{C}}$) is defined as the $L^2$-norm between the refined $\bm{\hat{S}}^{\mathrm {r}}_{i}$ and ground truth $\bm{S}_i$ segmentation.

In order to preserve the topology of the warped images, $L_{\mathrm{Reg}}$ is defined as:
$\lambda_{\mathrm{g}}\|\nabla \bm{U}_i \|_{2} + \lambda_{\mathrm{d}} \| \bm{U}_i \|_{2} + \lambda_{\mathrm{m}} \| \bm{\bar{U}}_i \|_{2},$
which regularizes the continuity and smoothness of the predicted deformation field \cite{evan2020learning}.
$\bm{U}_i$ represents the displacement field ($ \bm{\Phi}^{(t=1)} - \bm{\Phi}^{(t=0)}$), and $\lambda_{\mathrm{g}}$, $\lambda_{\mathrm{d}}$ and $\lambda_{\mathrm{m}}$ are the hyper-parameters for tuning the weight of the regularization loss. 
Finally, the overall loss $\mathcal{L}$ is the linear combination of all four losses: 
$\mathcal{L} = \mathcal{L}_{\mathrm{S}} + \mathcal{L}_{\mathrm{R}} + \mathcal{L}_{\mathrm{C}} + \mathcal{L}_{\mathrm{Reg}}.$


\section{Evaluation and Results}

\noindent\textbf{Data:} We train and validate our model on $274$ T2 fetal MRI scans ($253$ patients) from the Developing Human Connectome Project (dHCP)\footnote{http://www.developingconnectome.org/}. 
These images are randomly split into $202$ training images (from $186$ subjects), $18$ validation images (from $18$ subjects), and $54$ testing images (from $49$ subjects). Note that there is no overlap of any single subject in different subsets. The GA attribute in this dataset ranges from $20.6$ to $38.2$ weeks. 
In order to improve the performance of learning-based deformation, all the input images are affinely aligned to a coarse fetal atlas \cite{serag2012construction}. 

We use the revised segmentation results provided by the dHCP pipeline using DrawEM \cite{makropoulos2014automatic} as the ground truth for training and evaluation, from which the atlas used in DrawEM is advanced to a fetal atlas \cite{serag2012construction} to improve the segmentation performance. 
The ground truth label maps can be imperfect because of the fetal image artifacts, which is the main motivation of using a conditional atlas as an anatomical prior. 
The segmentation labels consist of nine classes, namely, cerebrospinal fluid (CSF), cortical gray matter (CGM), white matter (WM), outliers, ventricles, cerebellum, deep grey matter (DGM), brainstem, and hippocampus. 

Table \ref{tab1} shows the selected hyper-parameters used in the CAS-Net. 
All the experiments are conducted on a PC with an NVIDIA GTX $3080$ GPU. Our model took $2$ hours for $500$ training epochs (generating the conditional atlas), and took around $2$ seconds during inference per MRI (outputting the segmentation results), which is much faster comparing to traditional atlas generation and brain tissue segmentation.

\begin{table*}[t]
  \centering
  \caption{\footnotesize Hyper-parameters in CAS-Net.}
  \scriptsize
    \begin{adjustbox}{max width=\textwidth}
    \begin{tabular}{|c|c|c|c|c|c|c|c|c|}
\cline{1-9}
     Hyperparameters & $l,w,h$ & $c$ & $T$ &  $\lambda_{\mathrm{i}}$ & $\lambda_{\mathrm{l}}$ & $\lambda_{\mathrm{g}}$ & $\lambda_{\mathrm{d}}$  &  $\lambda_{\mathrm{m}}$ \\
     \cline{1-9}
     Value & 64 & 10 & 6 & 2 $<$200 epoch &  1 $<$200 epoch & 200 & 500 & 200\\
       &  &  &  & 1 $\ge$200 epoch & 2 $\ge$ 200 epoch &  &  & \\
    \cline{1-9}
    \end{tabular}%
    \end{adjustbox}
  \label{tab1} 
\end{table*}


\noindent\textbf{Evaluation:} 
A 3D-UNet is used as a baseline for evaluating the segmentation performance compared to different variants of the CAS-Net. 
In order to compensate for the class imbalance, the loss term for the 3D-UNet is re-scaled for the different tissues according to the average volume of each tissue.
Table \ref{tab} demonstrates the Dice Similarity Coefficient (DSC) scores used to evaluate the accuracy of the results. 
It also shows that the proposed CAS-Net achieves a higher overall average accuracy of $85.2\%$, compared to $70.1\%$ using a 3D-UNet.
Furthermore, CAS-Net significantly improves the accuracy of the small or complicated structures that are more likely to be affected by motion or partial volume artifacts.
Whereas intensity-based methods, e.g. 3D-UNet, may fail to segment such labels accurately from bad quality images.

\begin{table*}[t!]
  \centering
  \caption{\footnotesize The segmentation performance of the different variants of CAS-Net compared to a 3D-UNet baseline in terms of Dice similarity coefficient (\%). The metrics are presented in the format of mean and standard deviation (sd).}
  \scriptsize
    \begin{adjustbox}{max width=\textwidth}
    \begin{tabular}{lccccccccccc}
    \toprule
    Methods & CSF & CGM & WM & Outlier & Ventricles & Cerebellum & DGM & Brainstem & Hippocampus & Overall\\

        \cline{1-11}
        3D-UNet & 18.0& 80.3& 90.1& 66.0& \textbf{85.9} & 86.5& 80.0& 54.1& 70.0&   70.1\\
        (sd) & 2.3 & 2.8 & 3.2 & 7.4 & 3.5 & 3.9 & 2.2 & 5.0 & 6.5 & 4.1 \\
    
     	\cline{1-11}
        SS  & 86.7 &	83.2 &	91.9 &	\textbf{71.6} &	0 &		0 &		0 &		0 &		0 & 37.0\\
       (sd)  &	7.2 & 4.7 &	3.8 &	13.8 &	0 &		0 &		0 &		0 &		0  &  3.3\\
    	
    	\cline{1-11}								
        DRS & 84.7 & 83.4 & 91.8 & 66.0 & 76.0 & 90.2 & 90.5 & 89.6 & 77.0 & 83.2\\
        (sd) &	7.0 &	3.9 &	3.1 &	19.6 &	6.0 &	4.5 &	1.8 &	3.7 &	4.7 & 6.0 \\
    	
    	\cline{1-11}							
        CAS-Net & \textbf{87.7} &	\textbf{85.1} &	\textbf{92.9} &	71.5 &	78.3 &	\textbf{91.7} &	\textbf{91.1} &	\textbf{90.1} &	\textbf{78.5} & \textbf{85.2}\\
        (sd) & 7.2 &	4.1 &	2.8 &	13.5 &	5.7 &	4.0 &	2.0 &	2.8 &	4.6 & 5.2\\

    \toprule
    \end{tabular}%
    \end{adjustbox}

  \label{tab} 
\end{table*}


Table \ref{tab} also demonstrates that the segmentation results from using different variants of the proposed CAS-Net based on the selected sub-networks, namely, segmentation subnet (SS) and diffeomorphic registration subnet (DRS). 
Note that the SS is not supervised by the re-scaled loss, which results in a better segmentation performance for the salient tissue (larger volume) but fails to segment smaller tissues. 
Although the overall segmentation performance of the SS is inferior to the 3D UNet with a re-scaled loss, the main contribution of the SS within the CAS-Net architecture is to provide valuable information for the DRS to learn a good deformation field. 
The extension of the SS with the DRS significantly improves the overall accuracy to $83.2\%$ for all tissues, which can indicate that the predictions from the SS can provide valuable information for the DRS to learn the deformation field. 

At last, the final CAS-Net output is the combination of the SS and DRS, while SS learns to segment the input images based on solely the intensity values, and DRS learns to preserve the structural topology. The quantitative results from Tab.~\ref{tab} show that the performance from the final CAS-Net is better than both SS and DRS by $1.9\%$ and $1.7\%$, respectively, in terms of DSC for the CGM segmentation. Similar results can be found for the other tissues. 

\begin{figure}[htbp]
\centering
\hfill
\begin{subfigure}[b]{0.47\textwidth}
\centering
\includegraphics[width=1.0\linewidth]{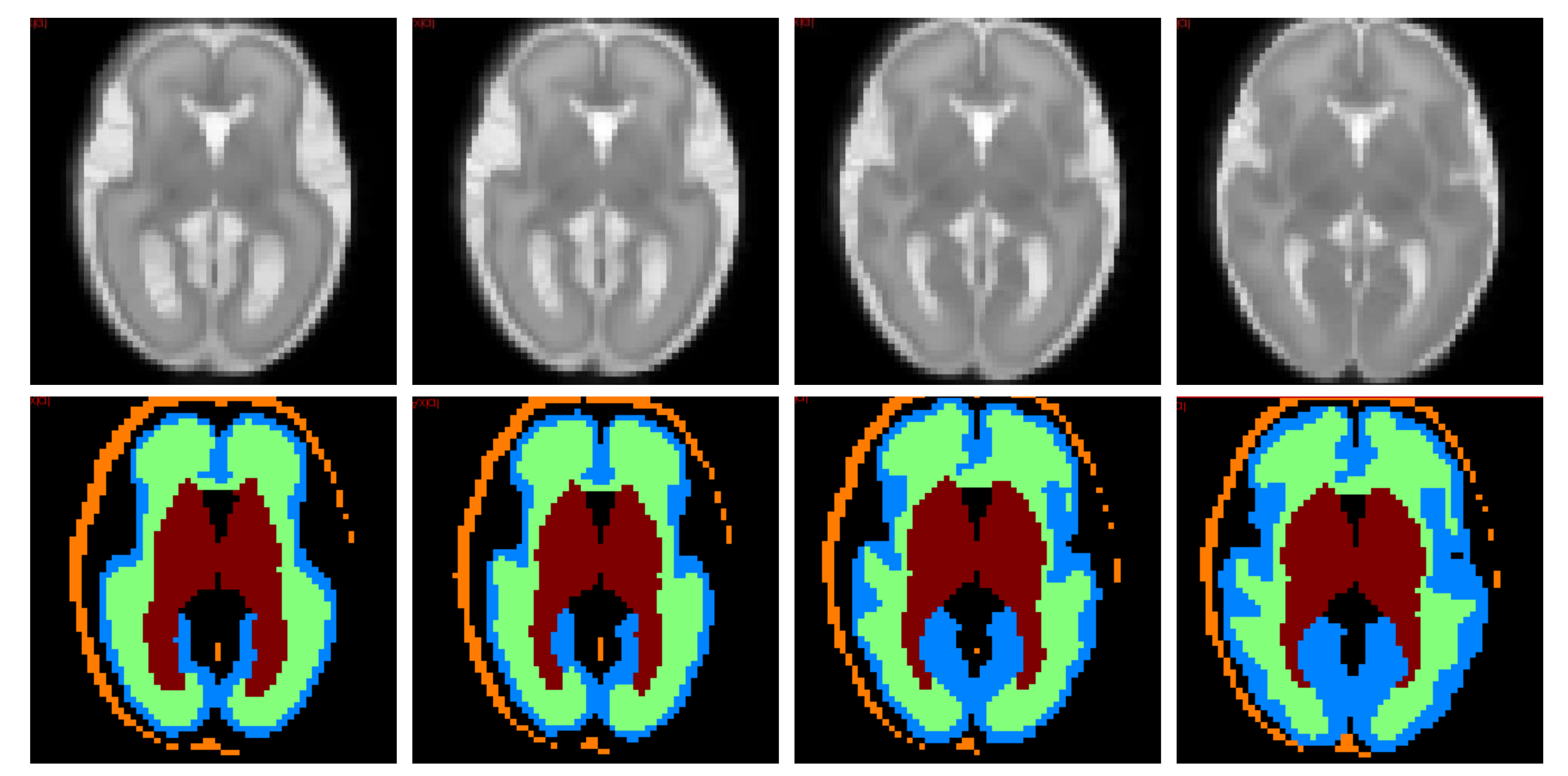}
\caption{  }
\label{fig:resultsA}
\end{subfigure}
\hfill
\begin{subfigure}[b]{0.5\textwidth}
\centering
\includegraphics[width=1.0\linewidth]{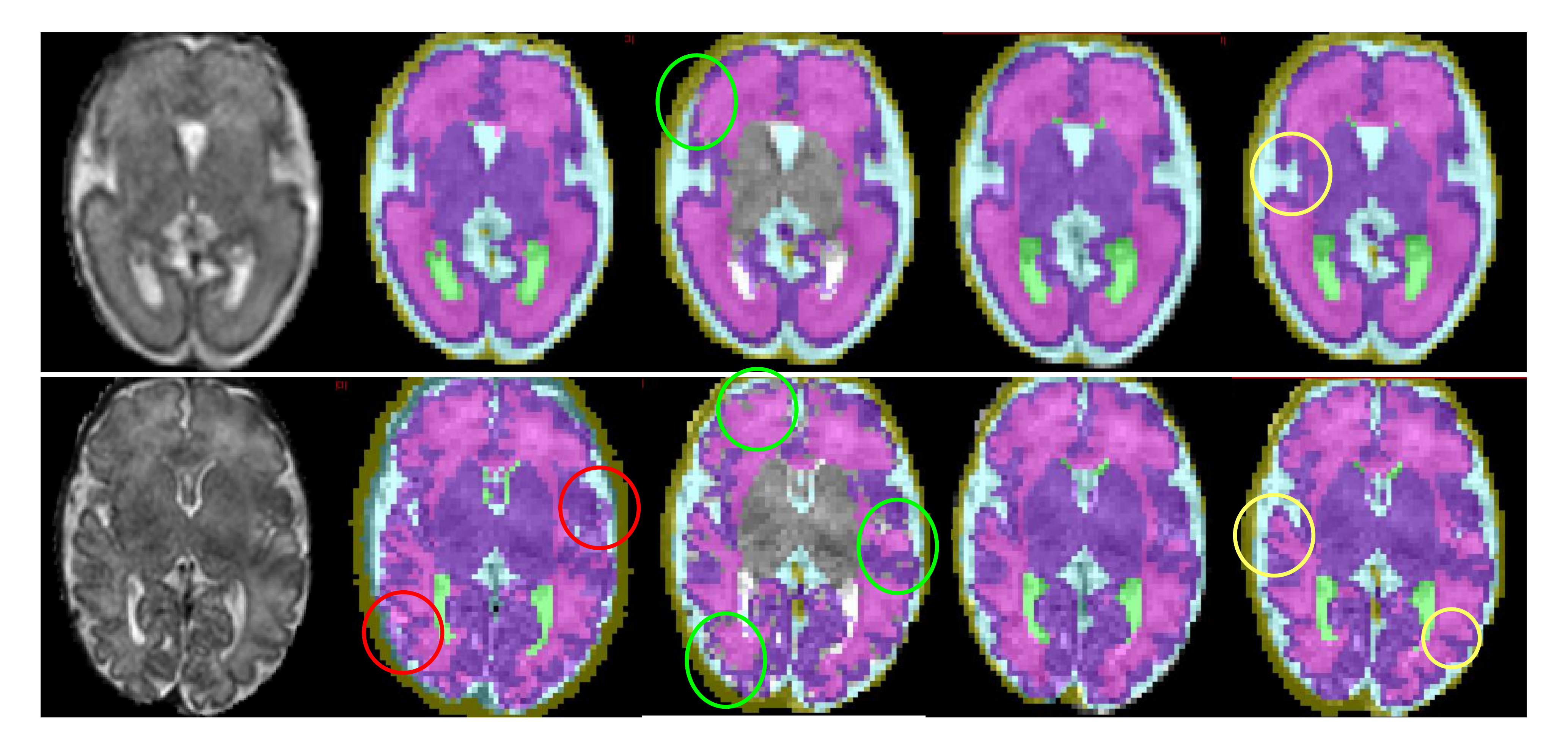}
\caption{ }
\label{fig:resultsB}
\end{subfigure}
\hfill
\caption{\footnotesize Generated conditional atlas image and labelmap (a): From left to right: lower than 25, 26-28, 29-32 and over than 33 weeks GA. Samples in the test set with imaging artifacts and their segmentation results from CAS-Net. (b): From left to right: original MRI in axial axis, groundtruth labelmap, segmentation results from SS, DRS, and combined layer, and their Dice score are 81.3, 82.5, 83.4 (upper for 27.6 GA subject) and 68.5, 71.2, 72.4 (lower for 37.0 GA subject), respectively. Note that the MRIs are shown in original resolution, 
and the segmentation maps are produced at lower resolution.}
\label{fig:results}
\end{figure}


\noindent\textbf{Discussion:} 
Figure \ref{fig:resultsA} shows that the complexity of the cortex of the generated conditional atlases increases with the GA. 
Furthermore, based on the conditional atlas, the segmentation maps from our model, including the output from the intermediate SS and DRS, as well as the final CAS-Net output, are shown in Fig.~\ref{fig:resultsB}. 
Compared with the SS output (the $3^{rd}$ column), the segmentation result from the DRS (the $4^{th}$ column), i.e., the atlas-based method, can preserve the connectivity for the CGM structure, while the discontinuous part from SS is highlighted by green circles. 
Comparing with the segmentation result between DRS and final CAS-Net, the final CAS-Net output can tightly follow the intensity changes of the tissues boundaries benefiting from the output from SS, which is highlighted by yellow circles. Consequently, our model combines the advantage from UNet-based (SS) and atlas-based (DRS) methods, which achieves better segmentation performance in terms of both connectivity and tissue accuracy. 
As shown in the second column, the ground truth we used to train the CAS-Net is not the golden labelmap that the segmentation errors are highlighted in red circle, but the output from our method can also correct this mislabelming, which indicates the potential for our pipeline to the noisy label problem.



\section{Conclusion}
Fetal MRI is an important tool to monitor the development of brain structure. In this paper, we first generate a model that learns a conditional atlas based on fetal MRI which shows the averaged morphological change across different age groups in the dHCP dataset. Given the conditional atlas with an anatomical prior, tissue segmentation performance improved with better mesh connectivity for the tissues when facing imaging artifacts. 
In future work, quantitative evaluation of the generated conditional atlas will be extended, and ablation studies for different subnets will be performed.

\noindent
\textbf{Acknowledgements:} Data in this work were provided by 
ERC Grant Agreement no. [319456]. We are grateful to the families who generously supported this trial.

\bibliographystyle{splncs04}
\bibliography{mybib}

\end{document}